\documentclass[preprint]{aastex}
\usepackage{psfig}

\newcommand{\citepeg}[1]{\citep[{e.g.,}][]{#1}}
\newcommand{\citepcf}[1]{\citep[{cf.}\phantom{}][]{#1}}

\def\etal{{\sl et al.}}
\def\lsim{\hbox{ \rlap{\raise 0.425ex\hbox{$<$}}\lower 0.65ex\hbox{$\sim$} }}
\def\gsim{\hbox{ \rlap{\raise 0.425ex\hbox{$>$}}\lower 0.65ex\hbox{$\sim$} }}

\def\f(h{\hbox{$~\!\!^{\rm h}$}}

\def\ale{\mathrel{\hbox{\rlap{\hbox{\lower4pt\hbox{$\sim$}}}\hbox{$<$}}}}
\def\age{\mathrel{\hbox{\rlap{\hbox{\lower4pt\hbox{$\sim$}}}\hbox{$>$}}}}

\slugcomment{Submitted to The Astronomical Journal, 5 December 2002}
\shortauthors{Bloom \etal~  }
\shorttitle{The small-scale redshift Distribution of GRBs}
\begin{document}

\title{Is the Redshift Clustering of Long-Duration Gamma-Ray 
Bursts Significant?}

\author{J. S. Bloom\altaffilmark{1,2}}

\bigskip 

\affil{$^1$ Harvard Society of Fellows, 78 Mount Auburn Street, 
Cambridge, MA 02138 USA}

\affil{$^2$ Harvard-Smithsonian Center for Astrophysics, MC 20, 
60 Garden Street, Cambridge, MA 02138, USA}

\begin{abstract}

The 26 long-duration gamma-ray bursts (GRBs) with known redshifts form
a distinct cosmological set, selected differently than other
cosmological probes such as quasars and galaxies. Since the
progenitors are now believed to be connected with active
star-formation and since burst emission penetrates dust, one hope is
that with a uniformly-selected sample, the large-scale redshift
distribution of GRBs can help constrain the star-formation history of
the Universe. However, we show that strong observational biases in
ground-based redshift discovery hamper a clean determination of the
large-scale GRB rate and hence the connection of GRBs to the star
formation history. We then focus on the properties of the small-scale
(clustering) distribution of GRB redshifts. When corrected for
heliocentric motion relative to the local Hubble flow, the observed
redshifts appear to show a propensity for clustering: eight of 26 GRBs
occurred within a recession velocity difference of 1000 km s$^{-1}$ of
another GRB.  That is, four pairs of GRBs occurred within 30
$h_{65}^{-1}$ Myr in cosmic time, despite being causally separated on
the sky. We investigate the significance of this clustering using a
simulation that accounts for at least some of the strong observational
and intrinsic biases in redshift discovery. Comparison of the numbers
of close redshift pairs expected from the simulation with that
observed shows no significant small-scale clustering excess in the
present sample; however, the four close pairs occur only in about
twenty percent of the simulated datasets (the precise significance of
the clustering is dependent upon the modeled biases). We conclude
with some impetuses and suggestions for future precise GRB redshift
measurements.

\end{abstract}


\section{Introduction}

A growing body of evidence suggests that long-duration GRBs arise from
the death of massive stars. If true, the large-scale redshift
distribution of GRBs should trace the star-formation history of the
Universe \citep{tot97,wbbn98,lr00,pm01,bl02}. And, as explosive and
transient events detectable to high redshift, GRBs clearly offer a
probe of the Universe that is distinct from other lighthouses
\citep{loeb02}. Already, GRBs have helped shed light on the nature of
damped Lyman $\alpha$ absorbers \citepeg{fmt+02} and the nature of the
faint end of the luminosity function of galaxies at moderate redshifts
\citepeg{dkb+01}.

As a redshift sample selected from an apparent isotropic population
and from hosts spanning the breadth of the galaxy luminosity function,
GRBs have the potential to provide a unique census of the redshift
distribution of matter in the Universe. Not only can the large-scale
distribution of GRBs help constrain the star-formation rate (SFR),
but, in a manner complementary to pencil-beam surveys and
magnitude-limited galaxy and quasar surveys, the small-scale
distribution can help test hypotheses and observational suggestions
about clustering and periodicity of sources in redshift space.

Given the penetrative powers of GRBs through dust, several studies have
attempted to compare the large-scale redshift distribution of GRBs
with the SFR obtained by other means
\citep{sdb01,lfr02,sts02,nor02}. Such studies focused on using the
prompt $\gamma$-ray properties of BATSE bursts, calibrated with some
measured GRB redshifts, to determine the bursting rate. With so few actual
redshifts measured, each of which were measured under different
observational constraints, the task of distinguishing different SFR
scenarios from the sparsely sampled GRB rate is considerably
challenging. Indeed we show herein, that when the observational biases
in redshift determination are taken into account, several proposed
forms for the universal SFR cannot be distinguished.

Investigation of the small-scale distribution of GRBs has not been
presented thus far. Of particular interest is whether the clustering
properties in redshift space are significant in light of the
observational biases and the intrinsic redshift distribution. Do GRBs
occur at preferred redshifts? There is enough historical, albeit
controversial, evidence to warrant an investigation of the
significance of special redshifts in the Universe using
GRBs. \citet{beks90}, for instance, found evidence for periodicity of
clustering of redshifts on 128 $h^{-1}$ Mpc scales in two pencil-beam
surveys. While excess power on these scales has not been definitively
confirmed in higher-redshift studies (e.g., with the 2dF redshift
survey: \citealt{hmm02}; although see \citealt{gd97} and
\citealt{dgn92}) the \citeauthor{beks90} result may be statistically
significant in comparison with $N$-body simulations
\citep{ycw+01}. But, as \citeauthor{ycw+01} emphasize, large clusters
in a galaxy sample tend to accentuate the appearance of excess power
at certain scales\footnotemark\footnotetext{Excess power on such 128
$h^{-1}$ Mpc scales could be an imprint of a primordial density
fluctuation \citep{dbps92}. As a statistical measure, if this
interpretation is true, then the same redshifts with over-densities
would not reoccur in surveys that sample different directions on the
sky. However, specific special redshifts would be preferred if the
observed power arises by more exotic means.  For example, oscillations
of the scalar potential \citep{mori91} or the gravitational constant
\citep{ssq96} induce an oscillation in the Hubble constant versus
cosmic time. The result is an apparent clustering of sources in
redshift space at epochs where the expansion temporarily slows.}.  A
GRB sample does not suffer such a bias since bursts occur in regions
of space that are gravitationally (and causally) disconnected.

The focus of this paper is on the small-scale (clustering)
distribution of GRB redshifts. In addition we examine the large-scale
distribution of GRB redshifts, and note the difficulty in determining
the global GRB rate due to the small sample and strong observational
biases.  In \S \ref{sec:sample} we present the GRB redshift sample for
26 bursts, corrected for the heliocentric motion through the local
Hubble flow and construct an observationally-motivated probability
distribution for redshift discovery. In \S \ref{sec:res}, we compare
the large-scale distribution of GRB redshifts with that expected from
the observing biases and an intrinsic rate distribution. After showing
the insensitivity of the present sample to distinguishing various
SFRs, and under the {\it anzatz} that GRBs {\it should} trace the SFR,
we fix a model for redshift discovery probability that adequately
reproduces the universal SFRs. Using this distribution, we then test
the significance of the number of observed pairs of GRBs with small
recession velocity differences. We conclude with a discussion about
the importance and usefulness of conducting precise redshift
measurements for future GRBs.

\section{The Redshift Sample}
\label{sec:sample}

Table \ref{tab:z} presents a summary of the sky location and observed
redshifts associated with 26 cosmological GRBs, in order of increasing
redshift. The highest observed redshift system derived using
absorption or emission features with the highest reported accuracy is
given in column seven. The line features used to measure the reported
redshift are given in column six. Absorption-line redshifts, which
place a strict lower limit to the actual GRB redshift, are found when
a spectrum of the afterglow is absorbed by metal-line systems along
the line of sight.  Emission-line redshifts are derived from
spectroscopy of the galaxy associated with the GRB position, either
found promptly, superimposed with the spectrum of the transient
afterglow (e.g., 980703; \citealt{dkb+98b}), or at later times, when
the afterglow light has faded. Nineteen sources in the sample have
measured emission-line redshifts, and 12 have absorption-line
redshifts. The uncertainty in the redshift measurements are usually
provided in the literature, but where none were provided we assumed an
error of 10 percent in the least significant digit reported.

Since we are interested in testing the observed rate and redshift
distribution of GRBs against different models for star-formation and
small-scale clustering, we now discuss the observational biases in GRB
detection, localization, and finally, redshift determination.

\subsection{Biases in GRB Detection}

If GRBs arise in collimated emission, then the relativistic Doppler
beaming allows for detection of a GRB only if the detector lies in the
direction of the cone of emission. By studying the opening angles of
those GRBs which have been detected, it is reckoned that only about
one in every five hundred GRBs are detectable at (i.e., pointed toward)
Earth due to beaming \citep{fks+01}.

Of those bursts that are beamed toward Earth, only those that reach a
certain critical flux level will be detected by the triggering
algorithms of a given satellite. The typical threshold for on-board
detection in the BATSE catalog is a peak flux of a 0.3 photon
cm$^{-2}$ s$^{-1}$ \citep{if94,phm98}. In finding that the total
energy release in GRBs is approximately constant, \citet{fks+01}
showed that bursts that are more highly collimated appear to be
intrinsically brighter (in flux/fluence per unit solid angle). Thus,
for a scenario where GRBs are jetted and release about the same amount
of electromagnetic energy, brighter bursts tend to be {\it more
distant}.  As a clear illustration of this, note that the peak flux of
the highest redshift burst, GRB\,000131 ($z = 4.5$), was at the
top 5\% of the entire BATSE catalogue \citep{ahp+00}.

This somewhat counterintuitive trend implies that the brightest GRBs
can be detected to extremely high redshifts, certainly beyond the age
of reionization ($z \sim 7$--10). More important for the present study is
that Malmquist bias should play little role in diminishing the observed
burst rate at high-redshift: bursts significantly fainter (by up to
$\sim10^{-3}$) than GRB\,000131 could have been detected at similar
redshifts.

Using the prompt high-energy properties, irrespective of individual
burst redshifts, it has been shown that the brightness distribution of
GRBs is consistent with a broad range of star-formation rates and
luminosity functions \citepeg{lw98,klk+00,sch01,sts02}. Using distance
indicators calibrated from a sub-set of the bursts with known
redshift, \citet{lrfrr02} claimed that the GRB rate increases
monotonically at least out to $z \approx 10$. In our opinion, this
suggestion points more to the pitfalls of extrapolating GRB distance
indicators beyond the redshift range in which they were calibrated,
than revealing concrete properties of the high-redshift bursting rate
density. Another approach to constrain the burst rate, which we
attempt herein, is to use only those bursts with actual measured
redshifts.

\subsection{Biases in Afterglow Localization}
\label{sec:darkbias}
For those GRBs that are triggered and rapidly localized using the
prompt X-rays or $\gamma$-rays, the observing community tries to determine a
sub-arcsec position and ultimately associate a redshift via host or
absorption-line spectroscopy. This is accomplished by first identifying a
transient afterglow.

Three major biases are important for afterglow discovery. First, GRBs
will be preferentially localized if they occur at a time and place in
the sky where ground- or space-based telescopes can rapidly observe
the prompt burst position with large enough fields--of--view. Second,
only GRB afterglows with brightnesses above a given threshold for the
particular afterglow-discovery observations will be localized. Last,
only GRB afterglows that are not heavily extincted by dust
obscuration will be first localized at optical wavelengths. Clearly,
radio and X-ray afterglow discoveries do not suffer this third
bias. The existence of a population of dark bursts \citep{dfk+01,pfg+02},
those bursts without detectable optical afterglow emission, is clearly
a result of all three biases. (The current dark burst debate resolves
around the relative importance of dust versus observational biases in
the manifestation of dark bursts \citep{bkb+02,fjg+01}.)

Since, for now, GRBs with redshifts are preferentially those that were
first localized at optical wavelengths (all but 970828, 980329, and
000210 were first localized at optical wavelengths), it seems
appropriate to suppose that a population of optically-localized GRBs
should trace a star-formation rate derived from other
optically-selected samples; \citet{pm01} have provided a parameterized
version of a SFR based upon the un-dust corrected Hubble Deep Field
measurements (model SF1); in this model the SFR drops beyond $z \sim
2$. In future GRB redshift samples of bursts that are first localized
at radio, X-ray or even infrared wavelengths, the expectation is that
the observed rate will more closely trace the true (high mass) SFR in
the universe.

\subsection{Biases in Redshifts Determination}

For those GRBs that are triggered and localized, most are followed up
spectroscopically. If an absorption or emission-line redshift is
found, it is of interest to know the relationship between this and the
true GRB redshift. A definitive measure of burst redshifts would need
to connect the bursting location with some stationary gas local to the
GRB, resulting in a detection of transient features. To date, no
transient absorption or emission features at optical wavelengths
features have been associated with a GRB afterglow. Transient
features, consistent with (later-time) optical spectroscopic redshift
measurements, have been seen in a prompt burst spectrum
\citep{afv+00,ldm+02} and a handful of X-ray afterglow spectra
\citepeg{pcf+99,pggs+00}. At present, the uncertainties in these
X-ray-derived redshifts are orders of magnitude larger than those
found typically at optical wavelengths. Perhaps more important,
redshifts derived purely from X-ray spectroscopy are contaminated from
an unknown (potentially relativistic) outflow speed of the emitting or
absorbing material; thus, precise measurements of the systematic
redshift of GRB progenitors are best determined using the UV and
optical features of host HII regions and/or intervening host clouds.

If a GRB occurs in intergalactic space, with no afterglow absorption
due to local gas, then the observed redshift would be systematically
lower than the true redshift. However, to date, the nature of the
highest $z$ absorption lines, suggests the presence of gas columns and
abundances not typically associated with tenuous outflow gas halos of
galaxies. Specifically, the equivalent widths of the highest redshift
absorption systems in GRB afterglows are significantly larger than in
systems seen though quasar lines--of--sight \citepeg{svk+02}. This is
naturally understood in the context of the prevailing progenitor
model, namely that long-duration bursts arise from the death of
massive stars and, owing to the rapid evolution of massive stars, GRB
explosion locations should be in or near relatively dense regions of
active star-formation. Thus, GRBs should reside in or near the region
that gives rise to the gas absorption.  Consistent with this picture
is the observation that well-localized GRBs, a subset of which
comprises the present sample, have been observationally connected to
the location of the light of the putative host galaxies \citep{bkd02}.

The probability of a spurious (spatial) association of an afterglow
with its putative host has been shown to be small ($\ale 10^{-2}$) for
most bursts. In fact in a sample of 20 GRBs with arcsecond
localizations, statistically, at most a few spatial associations could
have been spurious \citep{bkd02}. Therefore, from the location
argument, GRB redshifts determined from emission line spectroscopy of
the hosts are likely to be close to the true systematic velocities of
the progenitor; at most, we would expect a velocity offset on the
order of a few hundred km s$^{-1}$ from the motion of the progenitor
system about its host. Likewise, unless the afterglow light pierces
some high-velocity outflowing material, absorption-line redshifts
should also provide a accurate measure of the true GRB
redshift. Perhaps most convincing that measured redshifts are closely
related to the real systemic redshifts, is that of the five GRBs with both
emission and absorption line spectroscopy, the highest redshift
absorption system is always found to be consistent with the
emission-line redshift of the host.

\subsection{Correcting the measured redshifts for 
heliocentric motion relative to the local Hubble frame}
\label{sec:correctz}

One striking feature of the current redshift sample is the small
apparent redshift differences, on the order of a thousand km s$^{-1}$,
between some bursts from different sky locations and apparently random
trigger dates (differences of 6 months to 5 years). (Unlike in
pencil-beam surveys the proximity in redshift can have nothing to do
with systems, such as galaxies in clusters, that are in causal
contact.) These small velocity differences are comparable to the
velocity of the Solar System through the local Hubble frame, and so a
correction for this systematic motion is warranted. Such motion was
measured by the angular dependence of temperature variations in the
Cosmic Microwave Background (CMB) with the COBE satellite
\citep{sbk+91,fcc+94}. This measured velocity ($V_\odot = 365 \pm 18$
km s$^{-1}$) and direction (toward Galactic coordinates $l = 264.4 \pm
0.3$ deg, $b = 48.4 \pm 0.5$ deg;
\citealt{fcc+94}) are due to vector sum of the heliocentric motion
about the Galactic center, the peculiar motion of the Milky Way in the
Local Group and the Local Group in-fall. We assume that all reported
redshifts have been corrected to the heliocentric redshifts; this
correction is at most of order tens of km s$^{-1}$ and so this
assumption is relatively unimportant.

In order to remove this systematic bias from a measured redshift $z$,
we find the angle ($\theta_{\rm CMB}$) between the GRB position and
the heliocentric motion through the CMB; we then find the projected
heliocentric velocity toward the GRB as $v_{\odot, {\rm proj}} =
V_\odot \cos \theta_{\rm CMB}$. If $u c$ is the uncorrected apparent
recession velocity of a distant source, then, from the definition of
redshift and the formula for relativistic velocity addition, the
corrected redshift in the local Hubble frame (lhf) is found from:
\begin{equation}
(1 + z_{\rm lhf})^2 = 
  \frac
{1 + u v_{\odot, {\rm proj}}/c + u + v_{\odot, {\rm proj}}/c}
{1 + u v_{\odot, {\rm proj}}/c - u - v_{\odot, {\rm proj}}/c},
\label{eq:eq1}
\end{equation}
with the uncorrected apparent recession velocity as,
\begin{equation}
u = \frac{ (1+z)^2 - 1}{(1+z)^2 + 1}.
\end{equation}
The calculated values of $z_{\rm lhf}$ are given in column eight of Table
\ref{tab:z}. The associated uncertainties are derived by error analysis
of equation \ref{eq:eq1}, assuming that the errors on $V_\odot$ and
$\theta_{\rm CMB}$ are uncorrelated and noting that the fractional
error on $\theta_{\rm CMB}$ is significantly smaller than the
fractional error on $V_\odot$. As seen in Table \ref{tab:z}, for all
but the most accurately measured redshifts, this correction adds
negligibly to the fractional error of the redshift.

\section{Results}
\label{sec:res}

Table \ref{tab:z1} shows the absolute difference in redshift and
recession velocity of each burst matched with the burst that is
closest in apparent recession velocity. Figure \ref{fig:pairs} shows
this distribution of ``nearest neighbor'' apparent recession velocity
differences versus redshift. The smallest groupings occur around
redshift of unity, where the density of known GRB redshifts is
highest. Four pairs are closer than $\sim$1000 km s$^{-1}$
(0.0034\,$c$) in redshift space\footnotemark\footnotetext{Of the eight
bursts in the close four pairs, six redshifts were determined via
emission lines from the putative host galaxies. One redshift
(GRB\,000926) was one with absorption spectroscopy only and another
redshift (GRB\,000301C) was found with both emission and absorption
spectroscopy.}, and the largest velocity difference is between the
two highest redshift bursts ($|\Delta v| = 0.22\,c$).

Interestingly, the velocity differences of the four closest burst
pairs (at redshifts $z=$0.692, 0.844, 0.961, and $2.035$) are less
than the velocity dispersion of a large cluster of galaxies---and so,
given the unknown peculiar velocity of the GRB hosts and the
progenitor system within the hosts, these close burst pairs are
consistent with having occurred at the same cosmic time. Even assuming
no peculiar velocities of GRB progenitors relative to their local
Hubble flow, eight of 26 GRBs occurred within 30 $h_{65}^{-1}$ Myr of
another GRB.  This was determined by computing the look-back times
between the four close pairs and assuming a cosmology with $H_0 = 65$
km s$^{-1}$ Mpc$^{-1}$, $\Omega_{\Lambda} = 0.7$ and $\Omega_{\rm m}$
= 0.3. This cosmology is assumed throughout the paper where needed.

\subsection{Simplistic probability calculation}
\label{sec:simple}

How significant is this proximity in cosmic time? The bursts listed in
Table \ref{tab:z} have been detected with recession velocities in the
range $v_{\rm low} = 0.30\,c $ (GRB\,011121) to $v_{\rm high}$ = 0.94\,$c$
(GRB\,000131). In an ensemble of $n$ bursts, assuming uniform
probability of a GRB occurring and being detected between $v_{\rm
low}$ and $v_{\rm high}$, the probability that given pair of successive
bursts are not as close as a distance $\Delta v_{\rm c}$ is $P =
\exp\left(-n\, (v_{\rm high} - v_{\rm low})/\Delta v_{\rm c}\right)$.
Therefore, the probability that at least one close pair exists in the
ensemble, is,
\begin{equation}
P(1) = 1 - P^{n - 1}
\end{equation}
For $\Delta v_{\rm c} = 1000$ km s$^{-1}$ and $n = 26$, the
probability of at least one close pair with $\Delta v
\le 1000$ km s$^{-1}$ is 0.967. The probability of at least one close 
pair with at most the smallest velocity difference in the observed GRB
sample ($\Delta v_{\rm c} = 259$ km s$^{-1}$) is 0.585. Both these
probabilities thus indicate that there is nothing particularly unusual
about the occurrence of {\it at least one} close pair in the current
sample.

In general, the probability that exactly $k$ pairs are close can
be approximated as,
\begin{equation}
P(k) \approx \left[\frac{n\, (v_{\rm high} - v_{\rm low})}{\Delta v_{\rm c}}\right]^k
	\frac{P^{n - 1} \times n!}{(n - k)!\, k!}.
\label{eq:pk}
\end{equation}
For the present sample, there are $k = 4$ close pairs. Using this
simplistic calculation we therefore expect four pairs to occur by
random chance in 18.6\% of samples. We have verified this
approximation by a Monte Carlo simulation, selecting $n$ bursts
uniformly over the range [$v_{\rm low}, v_{\rm high}$]; however, this
approximation breaks down for large velocity differences as it does
not account for relativistic velocity subtraction between successive
bursts in velocity space.

\subsection{Probability calculation with a model distribution}

The above probability estimate does not take into account some
important observational and endemic biases. First, the rate of GRBs is
not uniform in redshift or recession velocity space. Instead, we
expect GRBs to trace (or at least approximate) the star-formation rate
\citep{tot97,wbbn98}, so that the peak of bursting activity should
occur around $z\sim 1-2$. This will make close velocity pairs more
likely at redshifts near unity. Second, the chance that a burst will
be {\it localized} well enough to follow-up with spectrometers is not
uniform in redshift. Instead, this chance depends, for optical
localizations, sensitively on the native dust obscuration of the
afterglow (see \S \ref{sec:darkbias}). Third, observing conditions and
instrumentation play a strong role in determining whether the
redshift of a GRB can be detected at that redshift. Most important,
emission lines and absorption lines falling outside the broadband
spectral coverage hamper the ability to detect redshifts (this may be
an explanation for why no redshifts have been detected for 980519,
980326, and 980329).  Even if a redshifted line falls within the
spectral range, the presence of night sky lines make detection of
redshifts at certain wavelengths more unlikely. Fourth, instrumental
sensitivity varies across the spectral range and from instrument to
instrument, night to night, and airmass to airmass.

Since we wish to know how the observed redshift sample compares with
that expected given the above considerations we construct a toy
probability model, $P(z)$, giving the differential probability that a
burst at redshift $z$ could ultimately yield a redshift
measurement. We construct a global (rather than individual)
distribution, neglecting the instrumental sensitivity differences from
burst to burst and the non-negligible chance that a redshift will be
incorrectly assigned even after the detection of one or more spectral
features. We use several parametrized versions of the star formation
rate (SFR) from \citet{pm01}.  The overall shape of the distribution,
$P_G(z)$, is then proportional to the number of bursts per unit
redshift per unit observer time; that is, $P_G(z)$ is proportional to
the co-moving volume element and the co-moving SFR ($\rho_{\rm SFR}$)
versus redshift:
\begin{eqnarray}
P_G(z) & \propto & \frac{dN}{dt\,dz} 
             \propto \frac{\rho_{\rm SFR}}{1 + z} \frac{dV}{dz} \label{eq:pgz}\cr
       & \phantom{=} & \propto \left(\frac{\exp(3.4\,z)}{\exp(3.8\,z) + 45}\right) \times \frac{d_L^2(z)} {(1+z)^3 \sqrt{\Omega_{\rm m}
(1+z)^3 + \Omega_{\rm k} (1+z)^2 + \Omega_\Lambda}},
\end{eqnarray}
following eqs.~[2]--[4] in \citet{pm01}. Here $d_L(z)$ is the
luminosity distance and $\Omega_{\rm m} + \Omega_{\rm k} +
\Omega_\Lambda = 1$. The factor of $(1+z)$ in the denominator accounts 
for the time dilation of the co-moving GRB rate. Following the discussion
in \S \ref{sec:darkbias}, we have nominally chosen the un-dust
corrected SFR from \citet{pm01} (SF1).  We also explore the two other
models for $P_G(z)$ in
\citet{pm01}.  In our model we do not take into account any biases 
related to the distance of the source in localization (in \S
\ref{sec:darkbias} we argued that there should be little bias in triggering
based on distance). However, because of the observed anti-correlation
between jet opening angles and prompt emission fluence \citep{fks+01},
and the theoretical correlation between prompt burst emission and
afterglow luminosity \citep{pk01a}, aside from dust obscuration, the
afterglow from {\it triggered} GRBs that originate from higher
redshift GRBs should not be significantly dimmer at the same observer
time (see also \citealt{cl00};
\citealt{lr00}).

Nevertheless, if there is any dependence of triggering/localization on
$d_L$, we can try to account for such using a probability function
related to luminosity distance, $P_L(z)$. There is also a clear effect
of distance upon redshift discovery: once a burst afterglow is
localized and a spectrum is acquired, the detectability of emission
features is diminished with increasing luminosity distance (nominally
as $d_L^{-2}$).  Systematically, only GRB hosts with higher rates of
unobscured star formation will be detectable at higher redshifts for a
given integration time. For a given afterglow brightness, the distance
bias does not exist for absorption-redshift GRBs as long as the
particular redshifted line is observable in the spectrum. To try to
account for these elusive effects, we take the probability of redshift
discovery due to distance [$P_L(z)$] as unity from $z = 0$ to $z_l$,
decreasing with $d_L$ to some power $L$. Nominally we take the values
of $z_l = 1$ and $L = -2$ but allow these quantity to vary in our
modeling (see \S \ref{sec:sfr}).  

Using the observability of emission and absorption lines in the
spectral range and in the presence of night skylines, we construct a
relative probability of redshift detection, $P_S(z)$. We take the
spectrum range as 3800--9800 \AA\ and assume a redshifted line is not
observable if it falls within a wavelength that is half the
instrumental resolution of a strong sky line listed in \citet{om92}.
Here, we assume the instrumental resolution of 5 \AA\ (dispersion of
$\sim$2.5\,\AA/pixel).  Based upon Table 1, we assume that a redshift
can be obtained unambiguously with the detection of at least one of
four star-formation emission features: Ly $\alpha$ $\lambda 1216$ \AA,
[O II] $\lambda\lambda 3727$ \AA, H$\alpha\,6563$
\AA, and [O III] $\lambda\lambda$ $4959,5008$ \AA. Also based on Table 1,
for redshifts based upon absorption features only, we require that at
least three of the following lines are detectable: Fe II
$\lambda$2344.2 \AA, Fe II $\lambda$ 2374.5 \AA, Fe II $\lambda$
2382.8 \AA, Fe II $\lambda$ 2586.7 \AA, Fe II $\lambda$ 2600 \AA, Mg
II $\lambda\lambda$ 2796.4, 2803.5 \AA, Mg I $\lambda$ 2853.0 \AA. If
no redshifted star formation line nor prominent absorption line is
observable at a given redshift, $z_0$, then we set $P_S(z_0) =
0$. Even if a line is observable, it may be too weak in emission or
too low in equivalent width to be detected. Therefore with less and
less lines observable at a given $z$, the value of $P_S(z)$ should be
diminished.

\begin{figure*}[tbp]
\centerline{\psfig{file=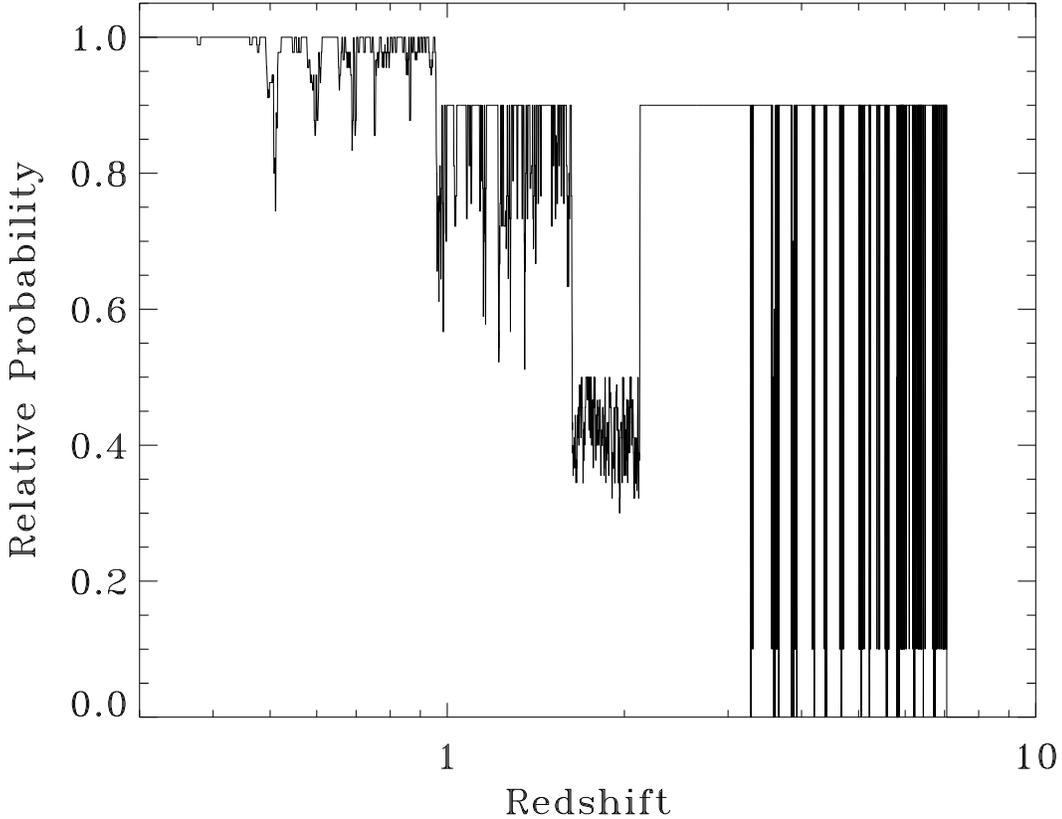,width=6.2in}}
\caption[]{A toy model for the redshift observability with 
ground-based optical spectroscopy in the presence of sky emission
lines, $P_S(z)$, for an instrumental resolution of 5 \AA.  The
observability of prominent star-formation emission lines (e.g., Ly
$\alpha$, [O II]) and gaseous halo absorption lines (e.g., Fe I, Mg
II) have been modeled using the wavelengths of bright night sky lines
as a mask; see text. The decrement of probability between $1.5 < z <
2$ is due to the inaccessibility of redshifted Ly$\alpha$, H$\alpha$,
and [O II] $\lambda\lambda$ 3727 \AA\ in the optical bandpass. No
optical redshift can be determined for bursts beyond $z\approx 7$, due
to Lyman $\alpha$ blanketing of the optical spectra. The assignment of
relative probability goes as follows: If one prominent emission line
is observable, then we set $P_S(z_0) = 0.9$. If more than 1 emission
line is observable then $P_S(z_0) = 1.0$. If no emission lines are
observable but at least a few prominent metal absorption lines are
observable, then $P_S(z_0) = 0.3$. If four of these nine absorption
lines are observable then $P_S(z_0) = 0.4$; for five observable lines,
$P_S(z_0) = 0.5$, etc. The main results of this paper are not strongly
sensitive to the precise values of these (arbitrary) probability
assignments.}
\label{fig:model2}
\end{figure*}

Figure \ref{fig:model2} shows our toy model for $P_S(z)$, based upon
the above considerations.  We stress that this is a toy model for
ground-based optical spectroscopy. Different instruments will provide
free-spectral ranges that differ from our nominal range (3800--9800
\AA). The resolution of each instrument also will also particularly
affect the observability of faint, narrow emission/absorption
features in the presence of sky lines. A moderate-resolution
spectrograph ($R \age 10000$) should allow for feature detection
closer in wavelength to a sky line; even features which partially
overlap a skyline could be detectable \citepeg{bbk+02}.

\begin{figure*}[tbp]
\centerline{\psfig{file=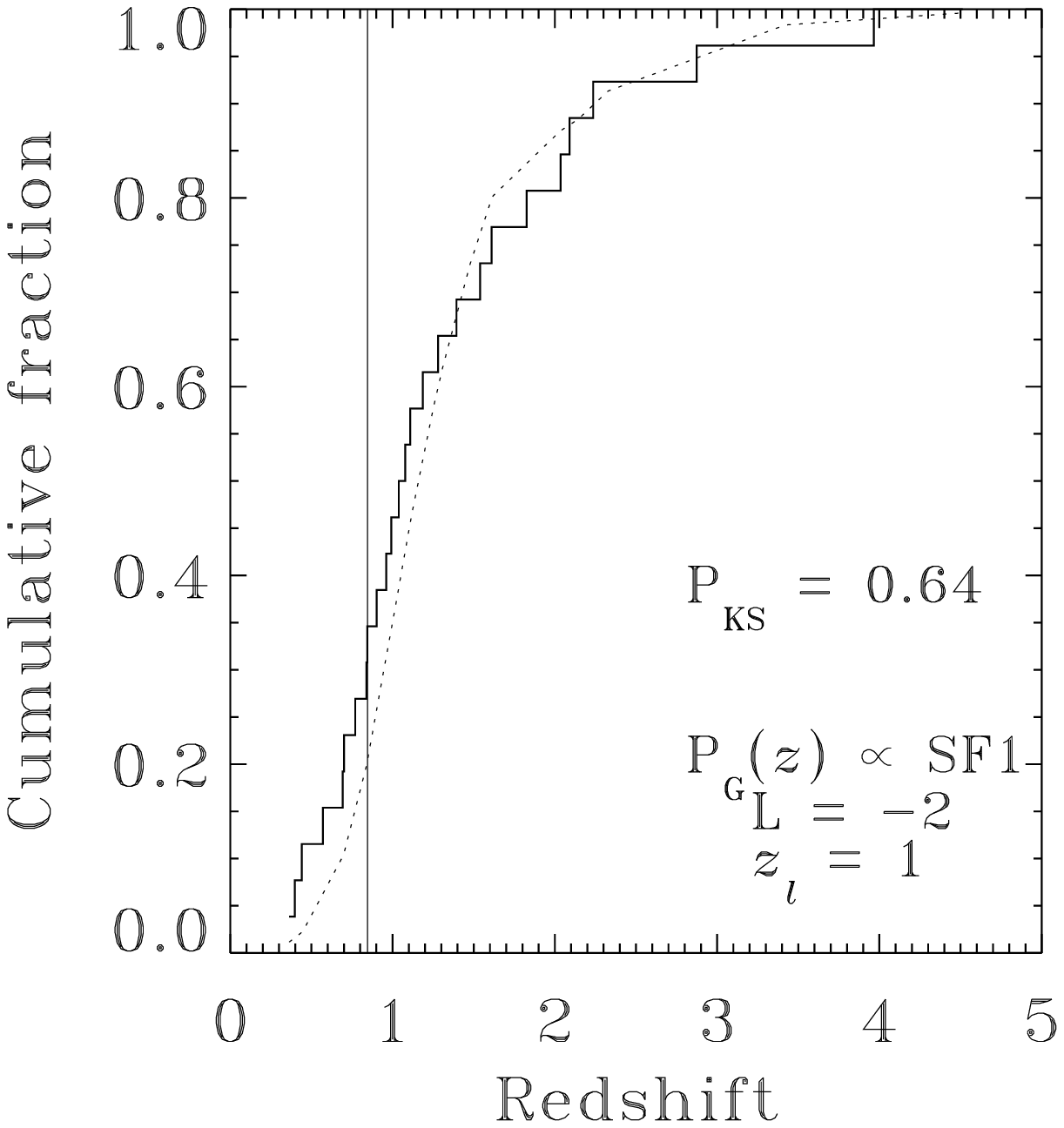,width=2.6in}
            \psfig{file=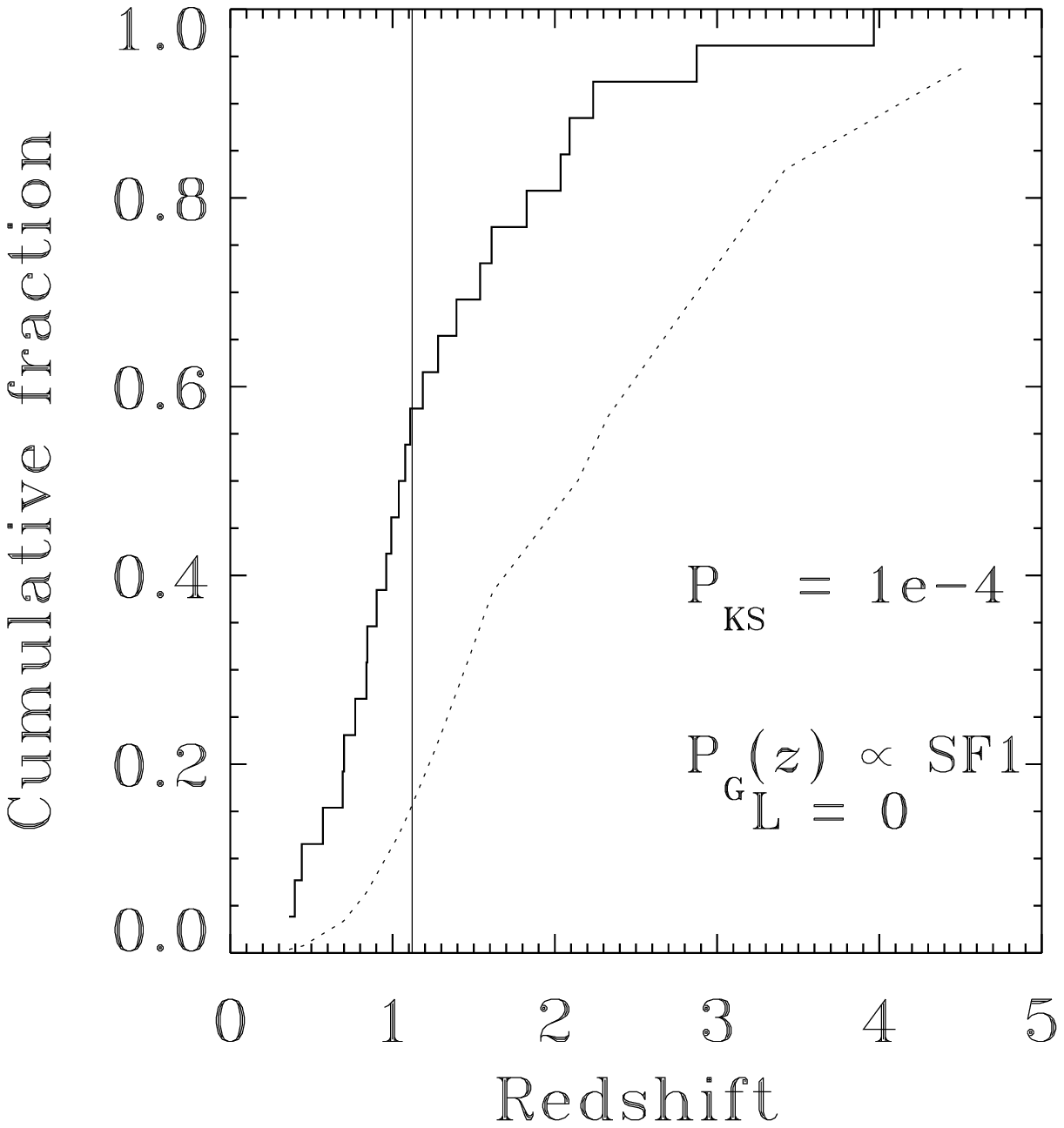,width=2.6in}}
\vskip -1cm
\centerline{\psfig{file=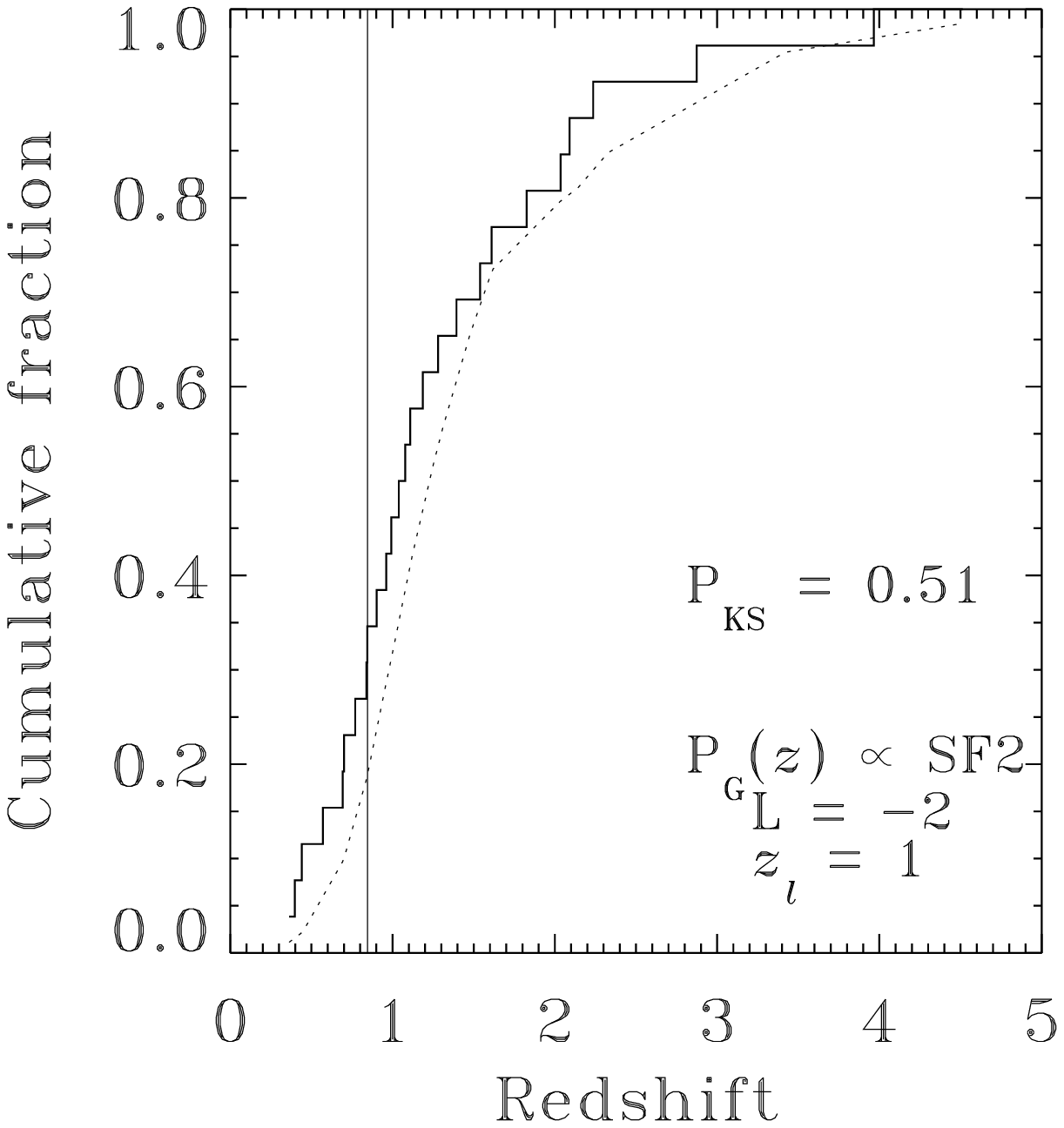,width=2.6in}
            \psfig{file=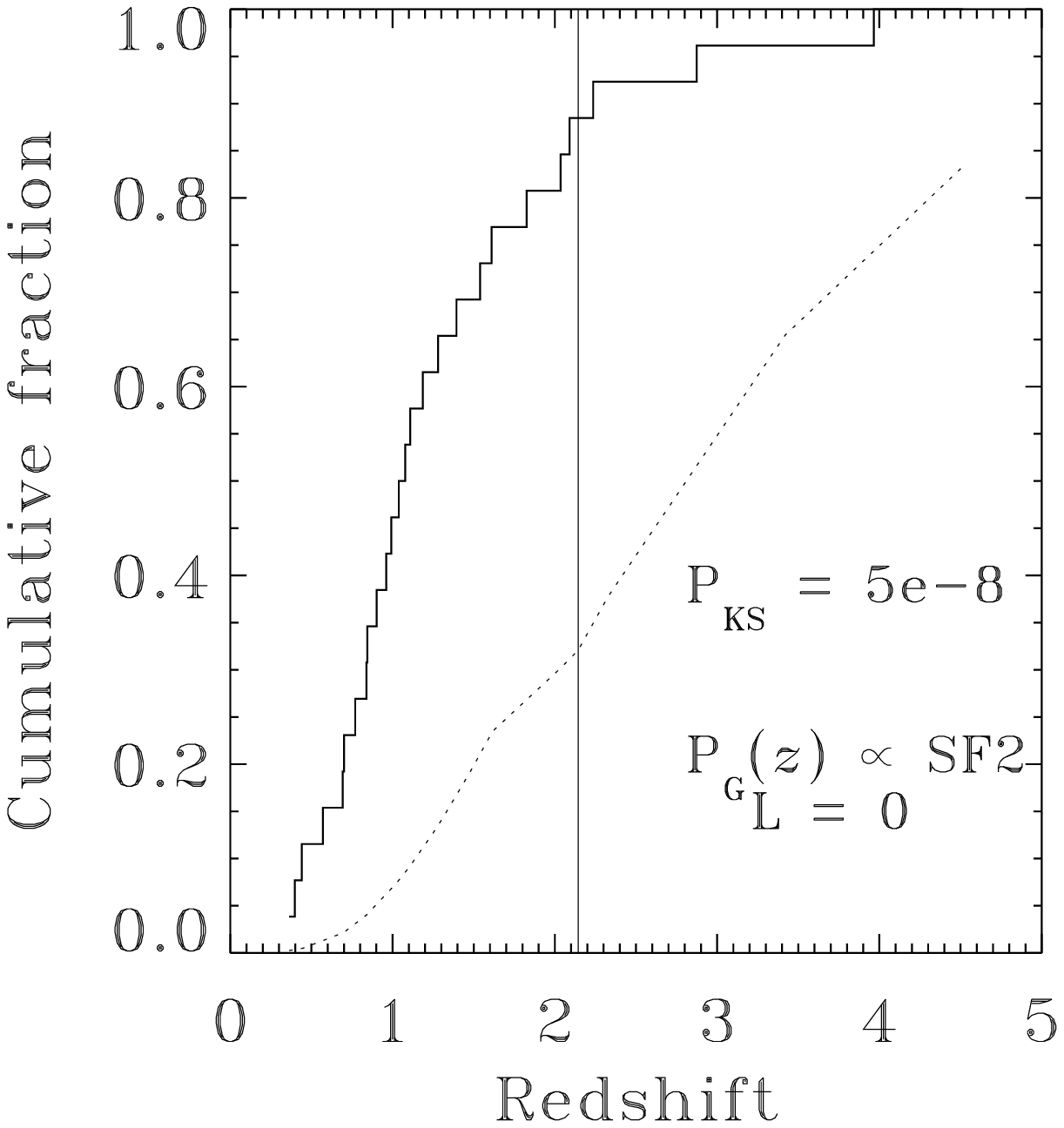,width=2.6in}}
\vskip -1cm
\centerline{\psfig{file=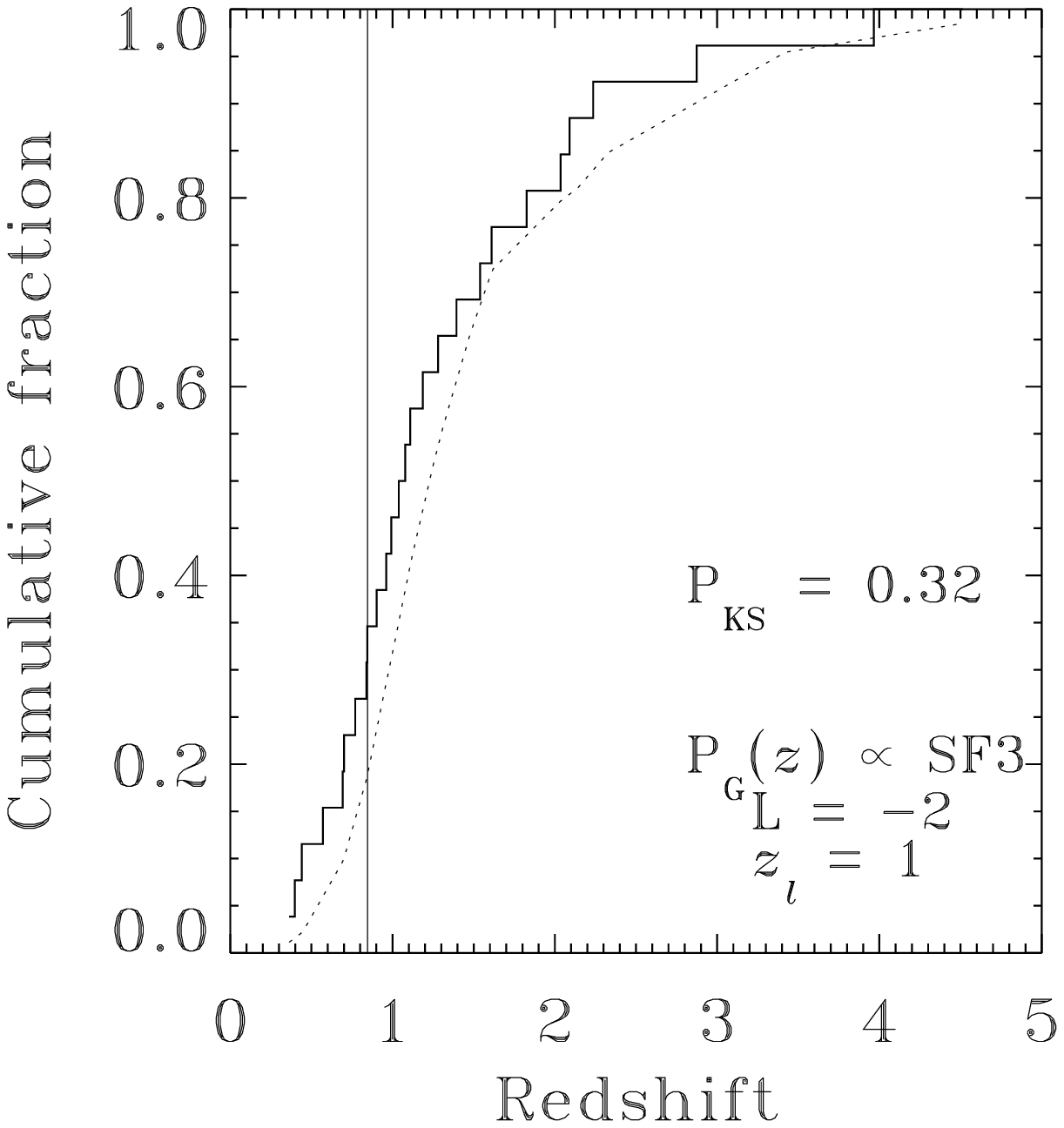,width=2.6in}
            \psfig{file=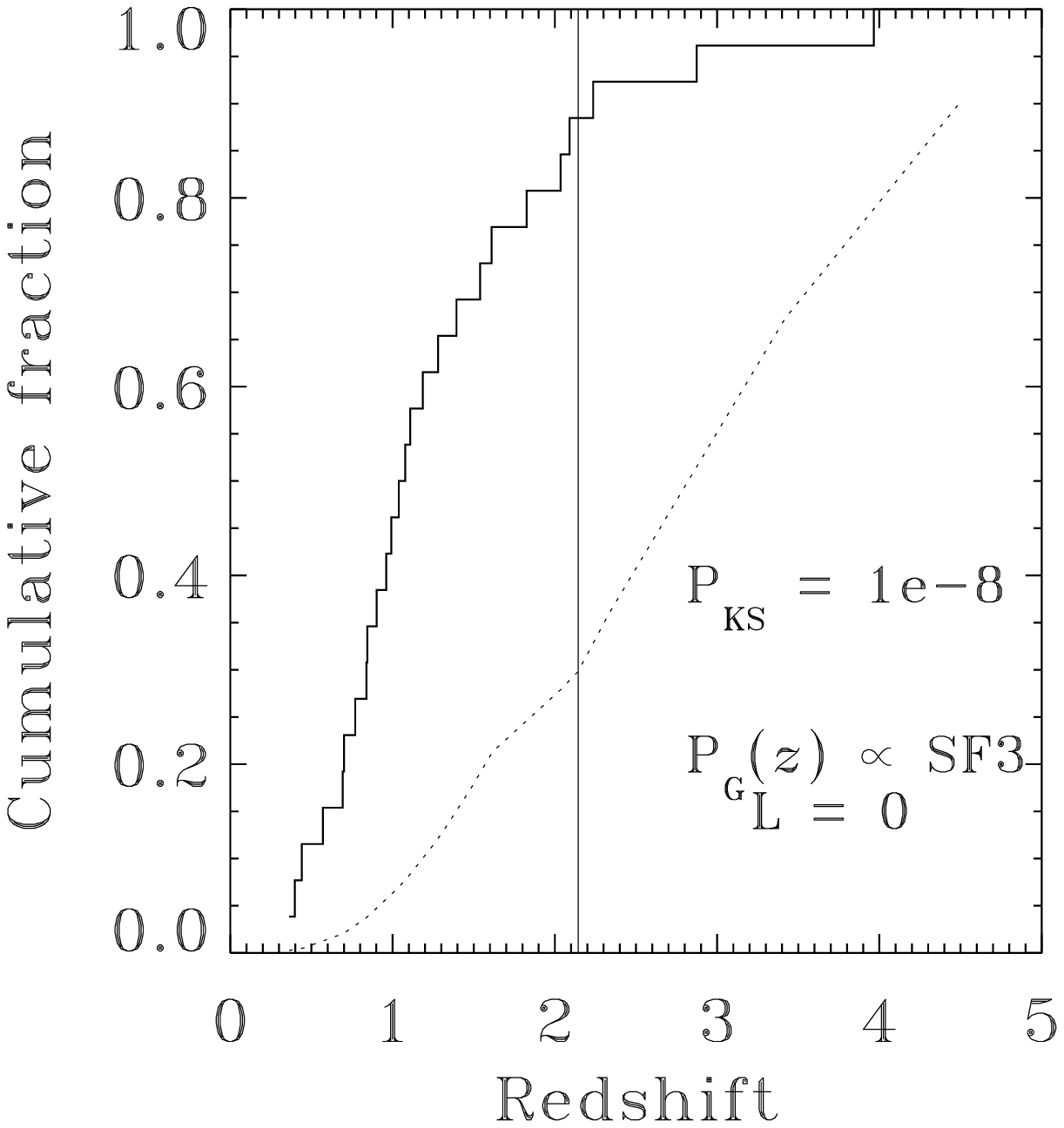,width=2.6in}}
\vskip -0.4cm
\caption[]{\small Comparison of the large-scale redshift distribution (solid cumulative line) with
various models for redshift detection. Models (shown as dotted lines)
that do not correct any observational biases in measuring high
redshifts ($L=0$) are clearly ruled out, but very different models for
the true rate, when including a simple model for the high redshift
bias ($L=-2$, $z_l \ale 1.25$), are all allowed by the data. The
Kolmogorov-Smirnov (KS) probabilities are given for six model
comparisons, with the redshift of maximum deviation from the model
noted with a vertical line. Clearly, the universal SFR cannot be {\it
inferred} from the current sample. Here we have used GRB rate models
that follow different parameterizations of the SFR (following
\citealt{pm01}). SF1 is the so-called Madau rate, where the GRB rate
falls beyond redshift of unity. SF2 is a dust-corrected form of the
Madau rate that levels off beyond redshift of unity
\citepeg{sag+99}. SF3 is a rate which continues to increase beyond $z
\approx 1.5$.}
\label{fig:kss}
\end{figure*}

\begin{figure*}[tbp]
\centerline{\psfig{file=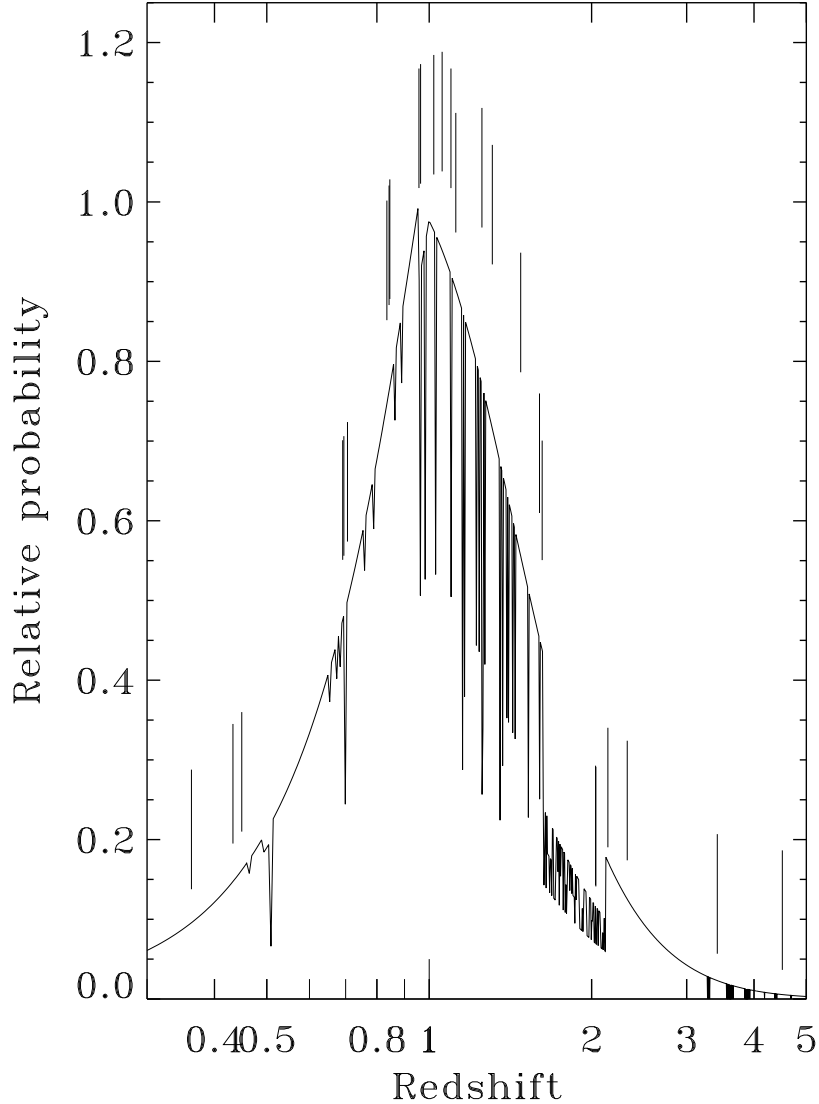,width=5in}}
\caption[]{A model distribution, $P(z)$, of the relative probability for
GRB redshift discovery using ground-based optical spectroscopy. The
model is shown with low resolution for clarity. The overall shape is
determined by the rate density, $P_G(z)$, constructed assuming that
the GRB rate should follow the Madau star-formation rate (SF1). Beyond
redshift of $z=1$ we assume that the relative detectability of
emission features in GRB hosts, drop as the inverse square of the
luminosity distance ($L=-2$). The uncertainty in the precise value of
$L$ (and hence $P_L(z)$) is also confounded by an uncertainty in the
correct form of the rate density. The toy model for the redshift
observability due to sky-lines is shown in Figure
\ref{fig:model2}.  The observed redshifts are noted as vertical lines
above the distribution. A machine-readable data file of the high
resolution version of this model may be obtained from the author by
request.}
\label{fig:model}
\end{figure*}

The resultant relative probability distribution of detecting an
optical redshift for a triggered GRB is constructed as $P(z) = P_G(z)
\times P_S(z) \times P_L(z)$ and is depicted in Figure
\ref{fig:model}.  The relative probability distribution was calculated in
bins of $\delta z = 0.001$.  As can be seen in the normalized
cumulative distributions shown in Figure \ref{fig:kss}, the overall
shape of the cumulative distributions are largely unaffected by the
sky lines. The most pronounced feature in $P(z)$ is the drop in
detection probability from $z \approx 1.5 - 2$, due to the
inaccessibility of strong emission lines in the optical bandpass. The
relative probability for detection of bursts at the redshift of
GRB\,980425 ($z=0.0088$) is exceedingly small and thus serves to
emphasize the distinction between GRB\,980425 and the other
long-duration GRBs with known redshift
\citepeg{gvv+98,bkf+98,schm00,pm01}. As such, we have not included
GRB\,980425 in our analysis.

\subsection{Consistency of the Large-Scale GRB Distribution with a Variety of SFR Models}
\label{sec:sfr}

As shown in Figure \ref{fig:kss}, the large-scale shape of the
observed redshift distribution is adequately described by the model
for $L=-2$, with the KS probability that the observed deviations are
consistent with a random selection from the model of $P_{\rm KS} =
0.64$ (SF1), $0.51$ (SF2), and $0.32$ (SF3). Models where no account
for the observational bias of detecting emission lines from
high-redshift GRB hosts ($L=0$) are clearly ruled out. While there are
small differences in the KS probability between various models for the
star-formation rate, such models cannot be statistically distinguished
by the current sample of GRBs with known redshifts. However, given the
low KS probabilities for $L=0$, the data do support the notion that
the observational biases of detecting emission lines from
high-redshift GRB hosts must be taken in account. Note that a fairly
large range of $L$ values ($\approx -1$ to $-3$ for SF1) yields a KS
consistency with the data.

How sensitive are these results to our toy model for $P_S(z)$ and
$P_L(z)$? Changing the value of $z_l$ we still get acceptable (but
lower) KS statistics for $L=-2$: $P_{\rm KS}$ = 0.20 ($z_l = 0.5$;
SF1), 0.22 ($z_l = 1.25$; SF1), 0.04 ($z_l = 1.5$; SF1). With $z_l
> 4.5$, the $L=-2$ case is effectively equivalent to the $L=0$ case
($P_{\rm KS} = 0.003$ for $z_l = 5$; SF1). By removing the effect of
the sky lines and limited free-spectral range in redshift
determination, the KS statistics are still acceptable, albeit with
lower values of the KS statistic: $P_{\rm KS} = 0.37$ (SF1), 0.19
(SF2), 0.10 (SF3).

\subsection{Testing the Small-Scale GRB Redshift Distribution}
\label{sec:small}

Assuming that GRBs {\it do} trace the global star-formation rate we
fix the form of $P_L(z)$ that gives a reasonable agreement with all
three SF models. From here, we can test the significance of the
small-scale clustering properties.  To do so, we produce a Monte Carlo
realization as sets of GRB redshifts drawn from the probability for
redshift detection $P(z)$ for each SFR model. We simulated 5000
iterations of sets of GRB redshifts. For each iteration, 26 bursts
were selected uniformly from [0,1] and then mapped to redshift using
the cumulative and normalized distribution of $P(z)$, interpolating
between bins to increase the resolution.

We thought of no obvious existing statistic to compare the small-scale
redshift structure of the observed distribution with the Monte Carlo
set.  However, since one feature is the existence of such close
redshift pairs we can ask how often such numbers of pairs are found in
the distribution.  Table \ref{tab:res} summarizes the results of the
comparison. Column two lists the number of observed GRBs within an
apparent recession velocity of $|\Delta v|$ of another GRB. Columns
three, four and five give the probability of such occurrences in the
simulated distributions for the three different GRB rates,
$P_G(z)$. 
 Following the table, the simulation predicts that at
least two bursts (one pair) should occur by random chance in 53\% of
real world samples for the smallest observed velocity difference. This
is close to the probability obtained in our simplistic calculation
(59\%) in \S \ref{sec:simple}.  For $\Delta v$ values near 1000 km
s$^{-1}$, the probability drops to $\approx$19\% before rising again
at large velocity offsets. Therefore, despite the apparent close
pairings in redshift space, we find no significant small-scale
redshift clustering in the present sample.

Note, however, this comparison only references 2-point
correlations. Following column eight of Table \ref{tab:z} and column 6
of Table \ref{tab:z1}, there are two groupings of three bursts within
2500 km s$^{-1}$ (at $z=0.69$ and $z=0.84$). Using the Monte Carlo
set, the probability of getting 2 groupings of three bursts within
2500 km s$^{-1}$ is 0.24 (SF1), 0.17 (SF2), and 0.13 (SF3). Again, the
present sample shows no evidence for significant multi-burst clustering.

\begin{figure*}[tbp]
\centerline{\psfig{file=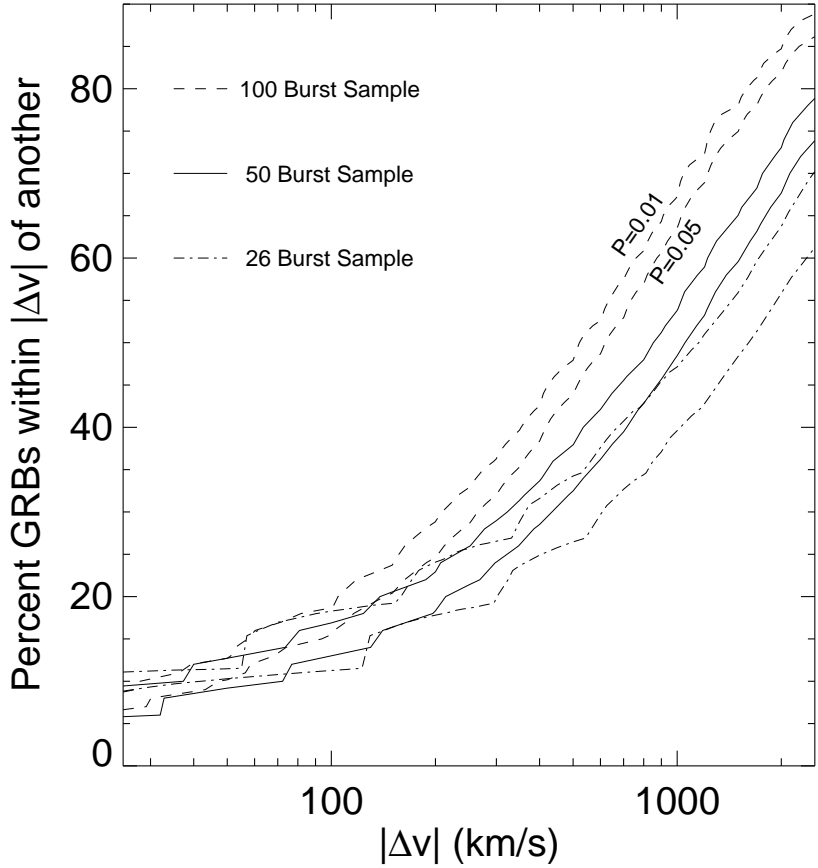,width=5in}}
\caption[]{Thresholds for the detection of significant small-scale 
clustering as a function of GRB redshift sample size. Based upon the
Monte Carlo simulation (with SF1), the curves show the observed
percentage of bursts required to be close to another burst as a
function of velocity difference for 95\% and 99\% confidence. For
instance in a sample of 100 ground-based optical redshifts, 66 bursts
must lie within 1000 km/s of another burst (i.e., 33 close pairs) to be
considered a statistically significant excess. Note that for all
samples sizes the percent requirements are similar for small velocity
offsets, but diverge at large velocity offsets: with more bursts, the
redshift density increases, so random pairing becomes more likely.}
\label{fig:future}
\end{figure*}

With the advent of {\it Swift}\footnotemark\footnotetext{{\tt
http://www.swift.psu.edu/}}, it is not unreasonable to expect upwards
of one hundred new GRB redshifts within the next several years. As the
density of redshifts increases, the number of close redshift pairs
(and triples) will certainly increase. Commensurately, for an expanded
sample to yield significant clustering results, the number of close
pairs must increase. Using the present models, as a function of bursts
in a sample, we predict the required number of observed close pairs in
order to be considered a significant excess above a random
sample. Figure \ref{fig:future} depicts the results. For the present
sample, approximately 1 (2) more close pairs would be needed with
velocity differences less than 1000 km s$^{-1}$, for the clustering
results to be considered significant at the 0.05 (0.01) level. From
the Figure, we see that in a sample of 50 GRB redshifts, there is a
5\% random chance of getting 5 close pairs within about 250 km
s$^{-1}$. If 6 or 7 more pairs are found, this would be considered a
significant result.

\section{Conclusions}
\label{sec:conc}

We have demonstrated, via a KS test, that the large-scale distribution
of observed long-duration GRB redshifts is compatible with having been
drawn from a reasonable model for the detection probability of burst
redshifts.  This supports the suggestion, based upon $\gamma$-ray
properties, that the rate of GRBs rises rapidly out to redshifts of
order unity \citepcf{schm00,lfr02,sts02}. To our knowledge, this is
the first investigation to claim this using only the observed GRB
redshifts and a model for the observational and intrinsic selection
function for GRB redshift discovery.

The data are relatively insensitive to various forms of the underlying
star-formation for a given set of redshift selection functions
$P_S(z)$ and $P_L(z)$. We have shown that by not considering the
biases of ground-based optical spectroscopy in redshift determination
(i.e., setting $P_S(z) = 1$) the large-scale redshift distributions
are still consistent with the SFR expectations, although at a lower
significance. (Without the presence of sky-lines, a derivation of the
functional form of $P_S(z)$ will be significantly more tractable with
redshifts obtained from space-based spectroscopy, such as from {\it
Swift}.) The large-scale rate results are much more sensitive to
$P_L(z)$, which is unfortunately the most difficult of the relevant
probabilities to determine {\it ab initio}; in general $P_L(z)$ should
be constructed on a case--by--case basis, including any biases of
distance upon triggering, localization probabilities, and redshift
determination.  There are certain psychological elements in the
redshift determination (e.g., integrating longer until a redshift is
found) that make this component particularly difficult to model. It is
clear, since the large-scale KS probability drops precipitously for
$z_l \age 1.25$, that there must be some strong diminution of the
detection rate of redshifts at larger distances.

Given these difficulties, since we cannot test which star-formation
rate GRBs trace best, we use the notion that GRBs {\it should} trace
the universal SFR as a point of departure. That is, we fix a
functional form of $P(z)$ that provides a reasonable agreement with
the expectations for the GRB rate on a large scale.  Under this
assumption ($L=-2$ and $z_l=1$), we test the significance of the
apparent small-scale clustering. As can be seen in Table
\ref{tab:res}, there does not appear to be any significant small-scale
clustering of GRBs in redshift space when compared with the Monte
Carlo set; however, the observed number of bursts paired with $|\Delta
v| \ale 1000$ km s$^{-1}$ occurs in only one out of five iterations.

The small-scale comparison in Table \ref{tab:res} is rather
conservative, however.  The only obvious {\it a posteriori} injection
to the comparison is in choosing the velocity bins based upon the
observed dataset. We have not required close pairs in the comparison
set to fall within a certain redshift range, say $z=[0.65,2.05]$, nor
have we required that the close pairs in the comparison sets be spaced
by at least $\delta z=0.1$ (as observed). Any such restrictions would
tend to increase the significance of the observed small-scale
clustering. However, without some {\it a priori} hypothesis that the
particular redshifts and spacings in the observed datasets are of
interest, further restrictions are unwarranted.

In the context of cosmologies with oscillating Hubble ``constants'' (see
\S 1) some {\it a priori} preference for special redshifts may indeed
exist. For example, we might expect that the most massive clusters of
galaxies would reside at redshifts where the Hubble constant loiters
at a local minimum. One of the most, if not the most massive clusters
known, with the highest observed X-ray temperature and high velocity
dispersion of member galaxies ($\sigma > 1100$ km s$^{-1}$), is
\hbox{MS 1054$-$03} \citep{na00,vdff+00}. Correcting the systemic
redshift reported in \citet{vdff+00} to the local Hubble frame
(following \S \ref{sec:correctz}), the redshift of the cluster is
$z_{\rm lhf} = 0.8337 \pm 0.0007$. This is just 177 $\pm$ 125 km
s$^{-1}$ from the redshift of GRB\,970508 yet the cluster and the GRB
site are separated by $\Delta \theta = 88.4^\circ$ on the sky.
Further, the redshift of \hbox{MS 1054$-$03} is 1456 and 1872 km
s$^{-1}$ from GRB\,990705 and GRB\,000210 (respectively), less than
two times the velocity dispersion of the cluster itself. Using our
Monte Carlo simulation with $P_G(z) \propto$ SF1, the probability that
three bursts out of 26 would fall within 2000 km s$^{-1}$ of
$z=0.8337$ by random chance is 0.035. Since the association with a
massive cluster was chosen {\it a posteriori} we do not claim this to
be a significant result, but it should be of interest to test the
association of new GRB redshifts with other massive clusters at
moderate to high redshifts.

The existence and expectation of several close groupings in redshift
space holds some practical implications for observations of future
GRBs. First, precise redshift determinations via moderate-resolution
optical spectroscopy ($R \age 1000$) will continue to be important
even though approximate redshifts, using GRB distance-indicators (such
as the Lag-Luminosity or Variability relations), may someday be used
reliably to constrain the overall large-scale redshift
distribution. Second, new future close redshift groupings will enable
efficient detailed narrow-band imaging studies of multiple GRB hosts,
such as the study of GRB\,000926 and GRB\,000301C \citep{fmt+02}.

\begin{figure*}[tbp]
\centerline{\psfig{file=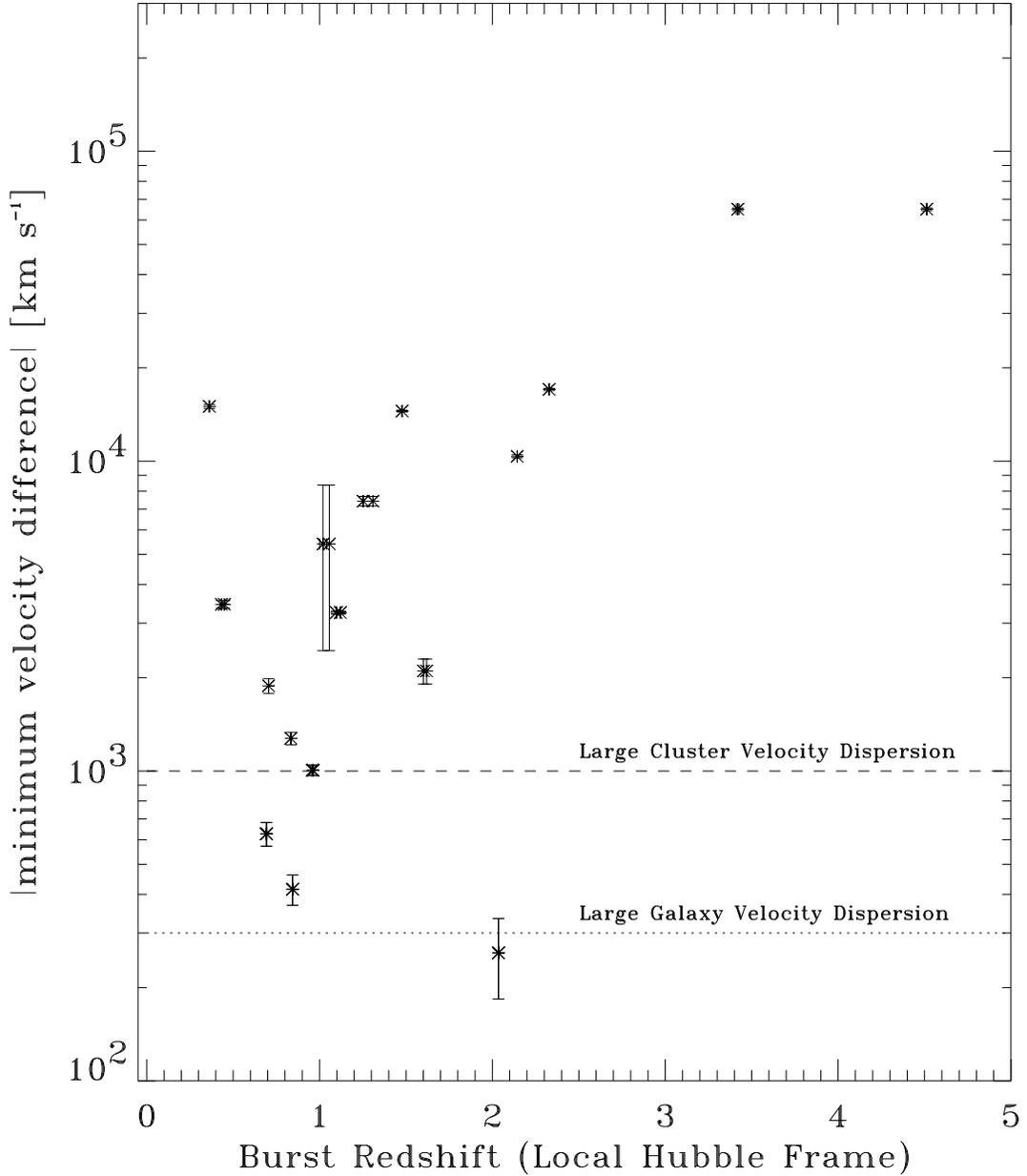,width=6.2in}}
\caption[]{The distribution of recession velocity differences between
nearest redshift pairs, following Table \ref{tab:z1}. The redshifts
have been correction for the heliocentric motion through the local
Hubble frame. \citealt{dgn92} performed a similar correction for
quasar redshifts but only accounted for the heliocentric motion in the
Galaxy. Representative velocity dispersions of galaxies and clusters
are indicated with the dashed and dotted lines.  There are 4 pairs of
bursts which are separated by $\ale 1000$ km s$^{-1}$.}
\label{fig:pairs}
\end{figure*}

We end by noting the curious absence of detected GRB redshifts in the
redshift range between $z \approx 2.3-3.2$, where a relatively clean
window for Lyman $\alpha$ emission detection exists at optical
wavelengths (see fig.~\ref{fig:pairs}). One explanation for this is
that redshift discovery via Ly $\alpha$ emission should be fruitless
for about half of high redshift GRBs, since the average equivalent
width of Ly $\alpha$ for Lyman Break Galaxies beyond $z \sim 3$ is
near zero (K.~Adelberger, personal communication); the absence of
bursts in this range could simply be due to the low number of GRBs
followed-up with sufficiently small delay to detect Lyman $\alpha$ in
absorption at optical wavelengths. In this respect, the uniformity and
sheer rate of redshift discovery with {\it Swift}, should be most
enlightening.

\acknowledgments

The author thanks D.~Frail and P.~van Dokkum for helpful suggestions
in improving the paper and gratefully acknowledges discussions with
K.~Adelberger at several stages of this project. The anonymous referee
is acknowledged and thanked for their insightful comments and
suggestions. The author is supported by a Junior Fellowship to the
Harvard Society of Fellows and by a generous research grant from the
Harvard-Smithsonian Center for Astrophysics.

\newpage

\newpage
\def\baselinestretch{0.9}

\def\gswetalag{1}
\def\vfketalaf{2}
\def\pkbetalag{3}
\def\pkbetalaga{4}
\def\bdkag{5}
\def\ddbetalae{6}
\def\bdkfad{7}
\def\ldmetalag{8}
\def\pfgetalag{9}
\def\dfketalaf{10}
\def\dkbetaladb{11}
\def\vrhetalae{12}
\def\pbketalag{13}
\def\dbkaf{14}
\def\bbketalag{15}
\def\bscetalag{16}
\def\bscetalag{16}
\def\cdketalag{17}
\def\kdoetalae{18}
\def\kdoetalae{18}
\def\cddetalaf{19}
\def\cdketalaf{20}
\def\hsgetalag{21}
\def\mgfetalag{22}
\def\kdretalad{23}
\def\ahpetalaf{24}
\begin{deluxetable}{lcccccrrl}
\tabletypesize{\scriptsize }
\tablecaption{Measured and corrected GRB redshifts\label{tab:z}}
\tablecolumns{9}
\tablehead{
\colhead{Burst} & \colhead{Pos.~(J2000)} & \colhead{$\theta_{\rm CMB}$} & \colhead{$v_{\odot,{\rm proj}}$} & \colhead{redshift} & \colhead{redshift} & \colhead{$z$} & \colhead{$z_{\rm lhf}$} & \colhead{Ref.} \\
\colhead{Name} & \colhead{$\alpha$/$\delta$ } & \colhead{$^{\circ}$} & \colhead{km s$^{-1}$} & \colhead{type} & \colhead{lines} & \colhead{} & \colhead{} & \colhead{} }
 \startdata 011121 & \parbox[t]{0.5in}{11:34:30.4 -76:01:41} & 69.1 &
 130.0 & em & \parbox[t]{1.0in}{{\tiny [O II], H$\beta$, [O III], He
 I, H$\alpha$}} & 0.362(1) & 0.363(1) & \gswetalag \\
 
990712
 & \parbox[t]{0.5in}{22:31:53.1 -73:24:28} &     99.4
 &    -59.4
 & em/abs & \parbox[t]{1.0in}{{\tiny [O II], [Ne III], H$\alpha$, H$\beta$, [O III]}}
 &   0.4331(2) &   0.4328(2) & \vfketalaf \\
 
010921
 & \parbox[t]{0.5in}{22:55:59.9 +40:55:53} &    145.8
 &   -302.0
 & em & \parbox[t]{1.0in}{{\tiny [O II], H$\beta$, [O III], H$\alpha$}}
 &   0.4509(4) &   0.4494(4) & \pkbetalag \\
 
020405
 & \parbox[t]{0.5in}{13:58:3.1 -31:22:22} &     45.6
 &    255.5
 & em & \parbox[t]{1.0in}{{\tiny [O II], [Ne III], H$\beta$, H$\gamma$, [O III]}}
 &  0.68986(4) &  0.69130(8) & \pkbetalaga \\
 
970228
 & \parbox[t]{0.5in}{05:01:46.7 +11:46:54} &     94.1
 &    -26.3
 & em & \parbox[t]{1.0in}{{\tiny [O II], [Ne III]}}
 &   0.6950(3) &   0.6949(3) & \bdkag \\
 
991208
 & \parbox[t]{0.5in}{16:33:53.5 +46:27:21} &     88.4
 &     10.2
 & em & \parbox[t]{1.0in}{{\tiny [O II], [O III]}}
 &   0.7055(5) &   0.7056(5) & \ddbetalae \\
 
970508
 & \parbox[t]{0.5in}{06:53:49.5 +79:16:20} &     92.3
 &    -14.7
 & em/abs & \parbox[t]{1.0in}{{\tiny [O II], [Ne III]}}
 &   0.8349(3) &   0.8348(3) & \bdkfad \\
 
990705
 & \parbox[t]{0.5in}{05:09:54.5 -72:07:53} &     83.6
 &     40.6
 & em & \parbox[t]{1.0in}{{\tiny [O II]}}
 &   0.8424(2) &   0.8426(2) & \ldmetalag \\
 
000210
 & \parbox[t]{0.5in}{01:59:15.6 -40:39:33} &    119.0
 &   -176.8
 & em & \parbox[t]{1.0in}{{\tiny [O II]}}
 &   0.8463(2) &   0.8452(2) & \pfgetalag \\
 
970828
 & \parbox[t]{0.5in}{18:08:34.2 +59:18:52} &    103.1
 &    -82.5
 & em & \parbox[t]{1.0in}{{\tiny [O II], [Ne III]}}
 &   0.9578(1) &   0.9573(1) & \dfketalaf \\
 
980703
 & \parbox[t]{0.5in}{23:59:6.7 +08:35:07} &    168.4
 &   -357.6
 & em/abs & \parbox[t]{1.0in}{{\tiny [O II], H$\delta$, H$\beta$, H$\gamma$, \hbox{[O III]}}}
 &   0.9662(2) &   0.9639(2) & \dkbetaladb \\
 
991216
 & \parbox[t]{0.5in}{05:09:31.2 +11:17:07} &     92.2
 &    -14.0
 & em/abs & \parbox[t]{1.0in}{{\tiny [O II], [Ne III]}}
 &     1.02(2) &     1.02(1) & \vrhetalae \\
 
000911
 & \parbox[t]{0.5in}{02:18:33.2 +07:45:48} &    134.0
 &   -253.5
 & em & \parbox[t]{1.0in}{{\tiny [O II]}}
 &   1.0585(1) &   1.0568(1) & \pbketalag \\
 
980613
 & \parbox[t]{0.5in}{10:17:57.6 +71:27:26} &     78.9
 &     70.0
 & em & \parbox[t]{1.0in}{{\tiny [O II], [Ne III]}}
 &   1.0969(2) &   1.0974(2) & \dbkaf \\
 
000418
 & \parbox[t]{0.5in}{12:25:19.3 +20:06:11} &     32.4
 &    308.2
 & em & \parbox[t]{1.0in}{{\tiny [O II], He I, [Ne III]}}
 &   1.1181(1) &   1.1203(1) & \bbketalag \\
 
020813
 & \parbox[t]{0.5in}{19:46:41.9 -19:36:05} &    122.7
 &   -197.3
 & abs & \parbox[t]{1.0in}{{\tiny Si II, \hbox{C IV}, \hbox{Fe II}, \hbox{Al II}, \hbox{Al III}, \hbox{Mg I}, \hbox{Mg II}, \hbox{Mn II}}}
 &    1.254(2) &    1.253(2) & \bscetalag \\
 
990506
 & \parbox[t]{0.5in}{11:54:50.1 -26:40:35} &     22.1
 &    338.2
 & em & \parbox[t]{1.0in}{{\tiny [O II] (resolved)}}
 & 1.306576(42) & 1.309180(135) & \bbketalag \\
 
010222
 & \parbox[t]{0.5in}{14:52:12.6 +43:01:06} &     70.4
 &    122.4
 & abs & \parbox[t]{1.0in}{{\tiny Si II, \hbox{C IV}, \hbox{Fe II}, \hbox{Al II}, \hbox{Sn II}, \hbox{Mg I}, \hbox{Mg II}, \hbox{Mn II}}}
 &   1.4768(2) &   1.4778(2) & \cdketalag \\
 
990123
 & \parbox[t]{0.5in}{15:25:30.3 +44:45:59} &     76.5
 &     85.1
 & abs & \parbox[t]{1.0in}{{\tiny Al III, Zn II, Cr II, \hbox{Zn II}, Fe II, Mg II, Mg I}}
 &   1.6004(8) &   1.6011(8) & \kdoetalae \\
 
990510
 & \parbox[t]{0.5in}{13:38:7.7 -80:29:49} &     75.4
 &     91.9
 & abs & \parbox[t]{1.0in}{{\tiny Al III, Cr II, Fe II, Mg II, Mg I}}
 &   1.6187(15) &   1.6195(15) & \vfketalaf \\
 
000301C
 & \parbox[t]{0.5in}{16:20:18.6 +29:26:36} &     82.1
 &     50.1
 & em/abs & \parbox[t]{1.0in}{{\tiny Fe II, Mg II, O I, \hbox{C II}, \hbox{Si IV}, Si II, C IV, \hbox{Fe II}, Al II}}
 &   2.0335(3) &   2.0340(3) & \cddetalaf \\
 
000926
 & \parbox[t]{0.5in}{17:04:9.8 +51:47:11} &     94.1
 &    -26.2
 & abs & \parbox[t]{1.0in}{{\tiny Si II, \hbox{C IV}, \hbox{Al II}, \hbox{Si II}, \hbox{Al III}, \hbox{Zn II}, \hbox{Fe II}, \hbox{Mg II}, \hbox{Mg I}}}
 &   2.0369(6) &   2.0366(7) & \cdketalaf \\
 
011211
 & \parbox[t]{0.5in}{11:15:18.0 -21:56:56} &     15.0
 &    352.6
 & abs & \parbox[t]{1.0in}{{\tiny Si II, Si IV, Cr II, Cr II, Si II, C IV, \hbox{Al II}, Fe III}}
 &    2.140(1) &    2.144(1) & \hsgetalag \\
 
021004
 & \parbox[t]{0.5in}{00:26:54.7 +18:55:41} &    158.4
 &   -339.4
 & em & \parbox[t]{1.0in}{{\tiny Ly$\alpha$}}
 &    2.332(1) &    2.328(1) & \mgfetalag \\
 
971214
 & \parbox[t]{0.5in}{11:56:26.0 +65:12:00} &     72.6
 &    109.1
 & em & \parbox[t]{1.0in}{{\tiny Ly$\alpha$}}
 &     3.42(1) &     3.42(1) & \kdretalad \\
 
000131
 & \parbox[t]{0.5in}{06:13:31.0 -51:56:40} &     75.2
 &     93.2
 & abs & \parbox[t]{1.0in}{{\tiny Lyman $\alpha$ forest}}
 &    4.511(2) &    4.513(2) & \ahpetalaf
\enddata

\tablecomments{Column (5) gives the method of redshift determination
for the burst (em = emission-line, abs = absorption-line) and column
(6) lists the specific atomic species used to measure the redshift
given in column (7). The redshift, corrected to the local Hubble
frame, is given in column (8), following from equation \ref{eq:eq1}.}
  
\tablerefs{
\gswetalag.~\citet{gsw+03};
\vfketalaf.~\citet{vfk+00};
\pkbetalag.~\citet{pkb+02};
\pkbetalaga.~\citet{pkb+02a};
\bdkag.~\citet{bdk01};
\ddbetalae.~\citet{ddb+99};
\bdkfad.~\citet{bdkf98};
\ldmetalag.~\citet{ldm+02};
\pfgetalag.~\citet{pfg+02};
\dfketalaf.~\citet{dfk+01};
\dkbetaladb.~\citet{dkb+98b};
\vrhetalae.~\citet{vrh+99};
\pbketalag.~\citet{pbk+02};
\dbkaf.~\citet{dbk00};
\bbketalag.~\citet{bbk+02};
\bscetalag.~\citet{bsc+03};
\cdketalag.~\citet{cdk+01};
\kdoetalae.~\citet{kdo+99};
\cddetalaf.~\citet{cdd+00};
\cdketalaf.~\citet{cdk+00};
\hsgetalag.~\citet{hsg+02};
\mgfetalag.~\citet{mgf+02};
\kdretalad.~\citet{kdr+98};
\ahpetalaf.~\citet{ahp+00}
}
\end{deluxetable}

\newpage

\begin{deluxetable}{lcccrr}
\tabletypesize{\small }
\tablewidth{5.2in}
\tablecaption{Nearest GRB neighbors in redshift and velocity space\label{tab:z1}
}
\tablecolumns{6}
\tablehead{
\colhead{Burst} & \colhead{$z_{\rm lhf}$} & \colhead{Nearest} & \colhead{$\Delta \Theta$} & \colhead{$| \Delta z |$} & \colhead{$| \Delta v |$} \\
\colhead{Name} & \colhead{} & \colhead{Burst Name} & \colhead{$^\circ$} & \colhead{} & \colhead{km s$^{-1}$} }
 \startdata
GRB 011121
 &    0.363(1) & GRB 990712 &     30.3 &    0.070(1) &        15053 $\pm$   223
 \\
 
GRB 990712
 &   0.4328(2) & GRB 010921 &    114.4 &   0.0166(4) &         3457 $\pm$    93
 \\
 
GRB 010921
 &   0.4494(4) & GRB 990712 &    114.4 &   0.0166(4) &         3457 $\pm$    93
 \\
 
GRB 020405
 &  0.69130(8) & GRB 970228 &    133.4 &   0.0036(3) &          628 $\pm$    55
 \\
 
GRB 970228
 &   0.6949(3) & GRB 020405 &    133.4 &   0.0036(3) &          628 $\pm$    55
 \\
 
GRB 991208
 &   0.7056(5) & GRB 970228 &    121.4 &   0.0107(5) &         1887 $\pm$   102
 \\
 
GRB 970508
 &   0.8348(3) & GRB 990705 &    152.1 &   0.0078(3) &         1278 $\pm$    59
 \\
 
GRB 990705
 &   0.8426(2) & GRB 000210 &     39.0 &   0.0026(2) &          416 $\pm$    47
 \\
 
GRB 000210
 &   0.8452(2) & GRB 990705 &     39.0 &   0.0026(2) &          416 $\pm$    47
 \\
 
GRB 970828
 &   0.9573(1) & GRB 980703 &     81.4 &   0.0066(2) &         1008 $\pm$    38
 \\
 
GRB 980703
 &   0.9639(2) & GRB 970828 &     81.4 &   0.0066(2) &         1008 $\pm$    38
 \\
 
GRB 991216
 &     1.02(1) & GRB 000911 &     42.3 &     0.04(1) &         5419 $\pm$  2967
 \\
 
GRB 000911
 &   1.0568(1) & GRB 991216 &     42.3 &     0.04(1) &         5419 $\pm$  2967
 \\
 
GRB 980613
 &   1.0974(2) & GRB 000418 &     54.6 &   0.0229(2) &         3253 $\pm$    35
 \\
 
GRB 000418
 &   1.1203(1) & GRB 980613 &     54.6 &   0.0229(2) &         3253 $\pm$    35
 \\
 
GRB 020813
 &    1.253(2) & GRB 990506 &    104.1 &    0.057(2) &         7446 $\pm$   266
 \\
 
GRB 990506
 & 1.309180(135) & GRB 020813 &    104.1 &    0.057(2) &         7446 $\pm$   266
 \\
 
GRB 010222
 &   1.4778(2) & GRB 990123 &      6.2 &   0.1233(8) &        14550 $\pm$    95
 \\
 
GRB 990123
 &   1.6011(8) & GRB 990510 &    126.2 &   0.0184(17) &         2109 $\pm$   195
 \\
 
GRB 990510
 &   1.6195(15) & GRB 990123 &    126.2 &   0.0184(17) &         2109 $\pm$   195
 \\
 
GRB 000301C
 &   2.0340(3) & GRB 000926 &     23.8 &   0.0026(7) &          259 $\pm$    75
 \\
 
GRB 000926
 &   2.0366(7) & GRB 000301C &     23.8 &   0.0026(7) &          259 $\pm$    75
 \\
 
GRB 011211
 &    2.144(1) & GRB 000926 &    105.4 &    0.107(1) &        10383 $\pm$   119
 \\
 
GRB 021004
 &    2.328(1) & GRB 011211 &    163.0 &    0.185(1) &        17082 $\pm$   132
 \\
 
GRB 971214
 &     3.42(1) & GRB 000131 &    134.1 &     1.09(1) &        65197 $\pm$   654
 \\
 
GRB 000131
 &    4.513(2) & GRB 971214 &    134.1 &     1.09(1) &        65197 $\pm$   654

\enddata
 
\end{deluxetable}

\begin{deluxetable}{lccccc}
\tablecaption{Testing the Significance of the Observed Small-scale Structure\label{tab:res}}
\tablewidth{5.3in}
\tablecolumns{6}
\tablehead{
\colhead{$|\Delta v|$} & \colhead{\# GRBs paired} &
\multicolumn{3}{c}{Prob.~(Observed$|$Simulation, $L=-2$)} & \colhead{Simple Prob.} \\
\colhead{km s$^{-1}$} & \colhead{$\le |\Delta v|$} & \colhead{$P_G(z) \propto$ SF1} & \colhead{$\propto$ SF2} & \colhead{$\propto$ SF3} &\colhead{}}
 \startdata
259 & 2   & 0.530 & 0.471 & 0.438 & 0.408 \\
416 & 4   & 0.303 & 0.247 & 0.215 & 0.252 \\
628 & 6   & 0.206 & 0.145 & 0.120 & 0.191 \\
1008 & 8  & 0.187 & 0.126 & 0.096 & 0.171 \\
1278 & 9  & 0.192 & 0.120 & 0.088 & 0.160 \\
1997 & 10 & 0.402 & 0.275 & 0.213 & 0.110 \\
2109 & 12 & 0.225 & 0.133 & 0.093 & 0.099 \\
3253 & 14 & 0.373 & 0.222 & 0.159 & 0.034 \\
3457 & 16 & 0.199 & 0.092 & 0.060 & 0.026 \\
5419 & 18 & 0.364 & 0.199 & 0.135 & 0.002 \\
7446 & 20 & 0.401 & 0.230 & 0.165 & 6e-5  \\
\enddata

\tablecomments{Columns (3)--(6) give the probability estimates of 
observing, by random chance, at least the observed number of GRBs
(column 2) that are paired with another within a velocity of $|\Delta
v|$ (column 1).  Columns (3)--(5) give the results from the Monte
Carlo simulations described in the text for different functional forms
of the universal star-formation rate. The last column gives the
simplistic probability calculation of observing {\it exactly} the
number of pairs assuming uniform selection in velocity space (equation
\ref{eq:pk} with $k$ = col.~[2]/2). The approximation breaks down for
large velocities because the probabilities of observing {\it more}
than the given number of bursts becomes non-negligible and
relativistic velocity subtraction is not taken in to account.}

\end{deluxetable}

\end{document}